%% file: Introduction.tex
\documentclass[11pt,fleqn]{book} 

\usepackage{tcolorbox}
\usepackage{wrapfig}
\usepackage{url}
\usepackage[utf8]{inputenc}
\usepackage{aas_macros}
\usepackage{pdfpages}
\usepackage{multicol}
\usepackage{setspace}
\usepackage{tcolorbox}

\input{structure}

\begin{document}
\pagenumbering{roman}

\includepdf{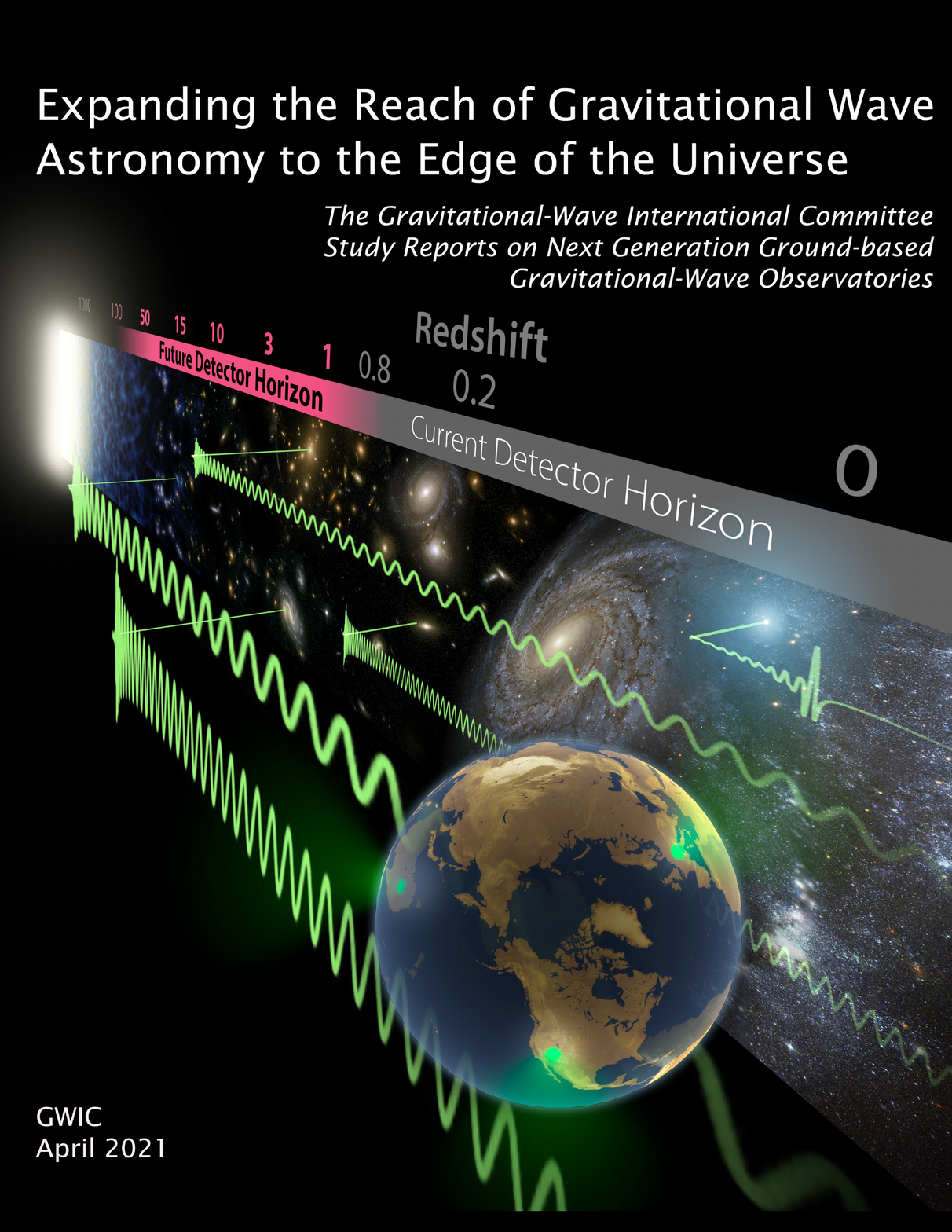}
\input{cover}


\pagenumbering{arabic}
\input{GWIC_Report_Intro} 
\clearpage

\end{document}

%% file: structure.tex

\usepackage[top=2.54cm,bottom=2.54cm,left=2.54cm,right=2.54cm,headsep=10pt,letterpaper]{geometry} 
\usepackage{xcolor} 
\definecolor{ocre}{RGB}{52,177,201} 

\usepackage{avant} 
\usepackage{mathptmx} 

\usepackage{microtype} 
\usepackage[utf8]{inputenc} 
\usepackage[T1]{fontenc} 

\usepackage{titlesec} 

\usepackage{graphicx} 
\graphicspath{{Pictures/}{figures/cosmo/}} 

\usepackage{lipsum} 

\usepackage{tikz} 

\usepackage[english]{babel} 

\usepackage{enumitem} 
\setlist{nolistsep} 

\usepackage{booktabs} 

\usepackage{eso-pic} 

\usepackage{titletoc} 

\contentsmargin{0cm} 
\titlecontents{chapter}[1.25cm] 
{\addvspace{15pt}\Large\sffamily\bfseries} 
{\color{ocre!60}\contentslabel[\LARGE\thecontentslabel]{1.25cm}\color{ocre}} 
{}
{\color{ocre!60}\normalsize\sffamily\bfseries\;\titlerule*[.5pc]{.}\;\thecontentspage} 
\titlecontents{section}[1.25cm] 
{\addvspace{5pt}\sffamily\bfseries} 
{\contentslabel[\thecontentslabel]{1.25cm}} 
{}
{\sffamily\hfill\color{black}\thecontentspage} 
[]
\titlecontents{subsection}[1.25cm] 
{\addvspace{1pt}\sffamily\small} 
{\contentslabel[\thecontentslabel]{1.25cm}} 
{}
{\sffamily\;\titlerule*[.5pc]{.}\;\thecontentspage} 
[]

\titlecontents{lsection}[0em] 
{\footnotesize\sffamily} 
{}
{}
{}

\titlecontents{lsubsection}[.5em] 
{\normalfont\footnotesize\sffamily} 
{}
{}
{}

\usepackage{fancyhdr} 

\fancypagestyle{plain}{\fancyhead{}} 

\makeatletter
\renewcommand{\cleardoublepage}{
\clearpage\ifodd\c@page\else
\hbox{}
\vspace*{\fill}
\thispagestyle{empty}
\fi}

\usepackage{amsmath,amsfonts,amssymb,amsthm} 

\newtheoremstyle{ocrenumbox}
{0pt}
{0pt}
{\normalfont}
{}
{\small\bf\sffamily\color{ocre}}
{\;}
{0.25em}
{\small\sffamily\color{ocre}\thmname{#1}\nobreakspace\thmnumber{\@ifnotempty{#1}{}\@upn{#2}}
\thmnote{\nobreakspace\the\thm@notefont\sffamily\bfseries\color{black}---\nobreakspace#3.}} 

\newtheoremstyle{blacknumex}
{5pt}
{5pt}
{\normalfont}
{} 
{\small\bf\sffamily}
{\;}
{0.25em}
{\small\sffamily{\tiny\ensuremath{\blacksquare}}\nobreakspace\thmname{#1}\nobreakspace\thmnumber{\@ifnotempty{#1}{}\@upn{#2}}
\thmnote{\nobreakspace\the\thm@notefont\sffamily\bfseries---\nobreakspace#3.}}

\newtheoremstyle{blacknumbox} 
{0pt}
{0pt}
{\normalfont}
{}
{\small\bf\sffamily}
{\;}
{0.25em}
{\small\sffamily\thmname{#1}\nobreakspace\thmnumber{\@ifnotempty{#1}{}\@upn{#2}}
\thmnote{\nobreakspace\the\thm@notefont\sffamily\bfseries---\nobreakspace#3.}}

\newtheoremstyle{ocrenum}
{5pt}
{5pt}
{\normalfont}
{}
{\small\bf\sffamily\color{ocre}}
{\;}
{0.25em}
{\small\sffamily\color{ocre}\thmname{#1}\nobreakspace\thmnumber{\@ifnotempty{#1}{}\@upn{#2}}
\thmnote{\nobreakspace\the\thm@notefont\sffamily\bfseries\color{black}---\nobreakspace#3.}} 
\makeatother

\newcounter{dummy}
\numberwithin{dummy}{section}
\theoremstyle{ocrenumbox}
\newtheorem{theoremeT}[dummy]{Theorem}

\newtheorem{exerciseT}{Exercise}[chapter]
\theoremstyle{blacknumex}
\newtheorem{exampleT}{Example}[chapter]
\theoremstyle{blacknumbox}

\newtheorem{definitionT}{Definition}[section]
\newtheorem{corollaryT}[dummy]{Corollary}
\theoremstyle{ocrenum}

\RequirePackage[framemethod=default]{mdframed} 

\newmdenv[skipabove=7pt,
skipbelow=7pt,
backgroundcolor=black!5,
linecolor=ocre,
innerleftmargin=5pt,
innerrightmargin=5pt,
innertopmargin=5pt,
leftmargin=0cm,
rightmargin=0cm,
innerbottommargin=5pt]{tBox}

\newmdenv[skipabove=7pt,
skipbelow=7pt,
rightline=false,
leftline=true,
topline=false,
bottomline=false,
backgroundcolor=ocre!10,
linecolor=ocre,
innerleftmargin=5pt,
innerrightmargin=5pt,
innertopmargin=5pt,
innerbottommargin=5pt,
leftmargin=0cm,
rightmargin=0cm,
linewidth=4pt]{eBox}	

\newmdenv[skipabove=7pt,
skipbelow=7pt,
rightline=false,
leftline=true,
topline=false,
bottomline=false,
linecolor=ocre,
innerleftmargin=5pt,
innerrightmargin=5pt,
innertopmargin=0pt,
leftmargin=0cm,
rightmargin=0cm,
linewidth=4pt,
innerbottommargin=0pt]{dBox}	

\newmdenv[skipabove=7pt,
skipbelow=7pt,
rightline=false,
leftline=true,
topline=false,
bottomline=false,
linecolor=gray,
backgroundcolor=black!5,
innerleftmargin=5pt,
innerrightmargin=5pt,
innertopmargin=5pt,
leftmargin=0cm,
rightmargin=0cm,
linewidth=4pt,
innerbottommargin=5pt]{cBox}

\makeatletter
\renewcommand{\@seccntformat}[1]{\llap{\textcolor{ocre}{\csname the#1\endcsname}\hspace{1em}}}
\renewcommand{\section}{\@startsection{section}{1}{\z@}
{-2ex \@plus -1ex \@minus -.2ex}
{1ex \@plus.1ex }
{\normalfont\large\sffamily\bfseries}}
\renewcommand{\subsection}{\@startsection {subsection}{2}{\z@}
{-2ex \@plus -0.1ex \@minus -.2ex}
{0.5ex \@plus.2ex }
{\normalfont\sffamily\bfseries}}
\renewcommand{\subsubsection}{\@startsection {subsubsection}{3}{\z@}
{-2ex \@plus -0.1ex \@minus -.2ex}
{.2ex \@plus.2ex }
{\normalfont\small\sffamily\bfseries}}
\renewcommand\paragraph{\@startsection{paragraph}{4}{\z@}
{-2ex \@plus-.2ex \@minus .2ex}
{.1ex}
{\normalfont\small\sffamily\bfseries}}

\usepackage[hidelinks,backref=page,pagebackref=true,hyperindex=true,colorlinks=false,breaklinks=true,urlcolor= ocre,bookmarks=true,bookmarksopen=false,pdftitle={Title},pdfauthor={Author}]{hyperref}

\newcommand{\thechapterimage}{}
\newcommand{\chapterimage}[1]{\renewcommand{\thechapterimage}{#1}}

\def\thechapter{\arabic{chapter}}
\def\@makechapterhead#1{
\thispagestyle{empty}
{\centering \normalfont\sffamily
\ifnum \c@secnumdepth >\m@ne
\if@mainmatter
\startcontents
\begin{tikzpicture}[remember picture,overlay]
\node at (current page.north west)
{\begin{tikzpicture}[remember picture,overlay]
\node[anchor=north west,inner sep=0pt] at (0,0) {\includegraphics[width=\paperwidth]{\thechapterimage}};
\draw[anchor=west] (2.46cm,-6cm) node [rounded corners=20pt,fill=ocre!10!white,text opacity=10,draw=ocre,draw opacity=1,line width=1.5pt,fill opacity=.6,inner sep=12pt]{\LARGE\sffamily\bfseries\textcolor{black}{\thechapter. #1\strut\makebox[22cm]{}}};
\end{tikzpicture}};
\end{tikzpicture}}
\par\vspace*{130\p@}
\fi
\fi}

\def\@makeschapterhead#1{
\thispagestyle{empty}
{\centering \normalfont\sffamily
\ifnum \c@secnumdepth >\m@ne
\if@mainmatter
\begin{tikzpicture}[remember picture,overlay]
\node at (current page.north west)
{\begin{tikzpicture}[remember picture,overlay]
\node[anchor=north west,inner sep=0pt] at (0,0) {\includegraphics[width=\paperwidth]{\thechapterimage}};
\draw[anchor=west] (2.46cm,-6cm) node [rounded corners=20pt,fill=ocre!10!white,fill opacity=.6,inner sep=12pt,text opacity=1,draw=ocre,draw opacity=1,line width=1.5pt]{\LARGE\sffamily\bfseries\textcolor{black}{#1\strut\makebox[22cm]{}}};
\end{tikzpicture}};
\end{tikzpicture}}
\par\vspace*{120\p@}
\fi
\fi
}
\makeatother

\usepackage{graphicx}
\graphicspath{{./figures/}}
\usepackage{appendix}
\usepackage{latexsym,amsmath,amsfonts,amssymb,booktabs}
\usepackage[font=small]{caption}
\usepackage{slashed,upgreek,amscd,cancel,tensor,color}
\usepackage{adjustbox}
\usepackage[numbers,sort&compress,square]{natbib}
\usepackage{epsfig,latexsym}
\usepackage{url}
\numberwithin{equation}{section}
\usepackage{doi}
\usepackage{subcaption}
\usepackage{mathtools}
\usepackage{upgreek}
\usepackage{stfloats}
\usepackage{afterpage}
\usepackage{multirow}

%% file: cover.tex

\begingroup
\thispagestyle{empty}



\newpage
\thispagestyle{empty}



\textcolor{white}{WHITE TEXT} 
\vspace{18cm}

\noindent \textsc{Gravitational Wave International Committee}\\

\noindent{This document was produced by the GWIC 3G Subcommittee}\\ 

\noindent \textit{Final release, April 2021}\\ 

\noindent \textit{Cover: Robert Hurt/IPAC/Caltech}


\chapterimage{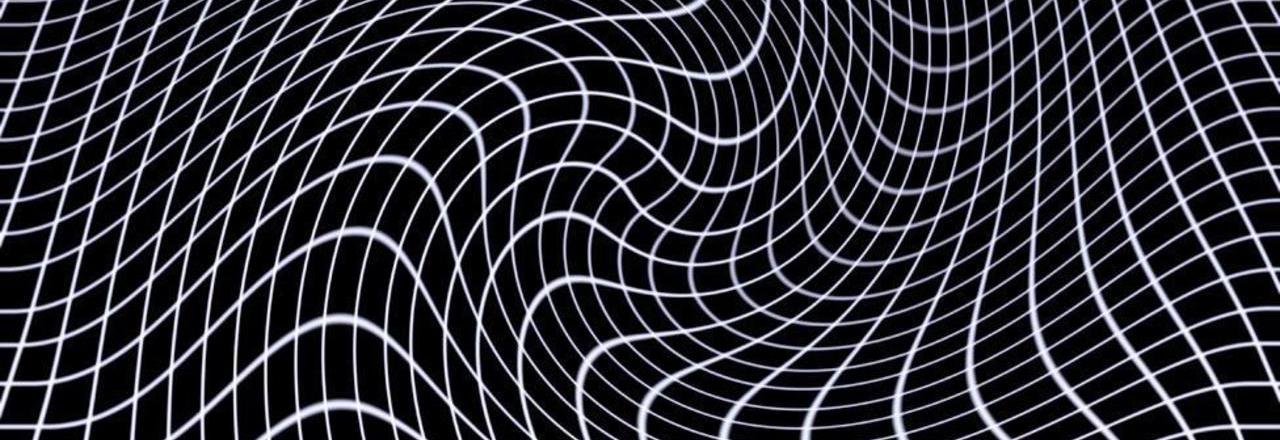} 
\pagestyle{empty} 
\tableofcontents 


\pagestyle{fancy} 
\newpage

%% file: GWIC_Report_Intro.tex
\chapterimage{Ultimate_Goals_pic.jpg} 
\chapter{Introduction}
\label{ch:introduction}

\vspace{0.5cm}
{\Large\bf T}he first direct detection of gravitational waves (GW) emitted from a pair of merging black holes located 1.3 billion light years from Earth in September 2015 has been heralded as one of most significant scientific breakthroughs in physics and astronomy of the 21st century.  Coming nearly 100 years after Albert Einstein predicted their existence, the first observations of directly detectable GW signal 
are ushering in a completely new way to observe and study the most violent and energetic astrophysical events in the Universe.

The opening of the GW window on the Universe has highlighted in dramatic fashion the tremendous scientific potential of this new field. The discoveries by LIGO and Virgo since the first detection have revealed stunning new insights into the nature of black holes and neutron stars.  

Motivated by these initial
breakthroughs and recognizing that to fully exploit the new field will require new observatories that may take 15 -- 20 years from conception until operations,the \href{https://gwic.ligo.org/index.html}{\textcolor{blue}{Gravitational Wave International Committee}} (GWIC)\footnote{The Gravitational Wave International Committee was formed in 1997 to facilitate international collaboration and cooperation in the construction, operation and use of the major gravitational wave detection facilities world-wide. It is associated with the International Union of Pure and Applied Physics as its Working Group WG.11. Through this association, GWIC is connected with the International Society on General Relativity and Gravitation (IUPAP's Affiliated Commission AC.2), its Commission C19 (Astrophysics), and another Working Group, the AstroParticle Physics International Committee (APPIC).} convened a subcommittee\footnote{The members of the GWIC 3G Subcommittee are provided in Section~\ref{3gteam}.} in 2016 to examine the path to build and operate a network of future ground-based observatories, capable of extending the observational GW horizon well beyond that currently attainable with the current generation of detectors. 

The primary aims of the \href{https://gwic.ligo.org/3Gsubcomm/}{\textcolor{blue}{Third Generation (3G) Subcommittee}} were to identify and frame the major issues and challenges in developing a coherent vision for future worldwide network of GW observatories and, from that, produce a resource for the ground-based GW community that can assist in planning and executing a strategy to realize the vision.  Over the period from November 2016 through March 2019, members of the subcommittee drew in researchers from the GW and broader research communities to collect input and produce reports in five major areas as described below. Each report was subsequently reviewed independently and anonymously by a panel comprised of funding agency officers, discipline experts and generalist scientists appointed by the \href{https://www.nsf.gov/mps/phy/gwac.jsp}{\textcolor{blue}{Gravitational Wave Agency Correspondents}} (GWAC).\footnote{The Gravitational Wave Agencies Correspondents was formed in 2015 to provide a direct channel of communication between funding agencies to coordinate the use of existing and explore new funding opportunities for the gravitational wave science community.}  The final reports were approved for release by the GWIC membership in September 2020.

\section{The GWIC 3G Subcommittee Reports}

The five reports commissioned by GWIC lay out an ambitious set of science targets achievable by future ground-based detectors, identify the requisite research end development in detector development and large-scale computing needed to reach those targets, examine the synergies and complementarity with other scientific disciplines, and assess possible governance models to manage and operate a unified global network of future observatories:\\


\begin{enumerate}
\item{\bf{The Next Generation Global Gravitational Wave Observatory Science Book}} (59 pages) presents a comprehensive survey of scientific frontiers accessible and the fundamental questions addressable by the future observatories, exploring the fields of high energy astrophysics, cosmology, gravitational physics, nuclear physics, and high energy physics. The report highlights a large number of key science targets available to future observatories possessing ten times the sensitivities attainable with today's LIGO, Virgo, and KAGRA observatories.  

\item \textbf{Research and Development for the Next Generation of Ground-Based Gravitational-Wave Detectors} (77 pages) examines in detail the wide range of nearer- and longer-term detector R\&D programs needed for next generation GW detectors commensurate with the key science targets presented in the Science Book, including considerations of site selection and large-scale vacuum infrastructure. The report makes a series of detailed recommendations on the needed advances in detector technology and the timescales needed to achieve those advances. It also identifies areas where larger-scale globally coordinated R\&D efforts will be critical to ensuring success while minimizing costs.   

\item \textbf{Future Ground-based Gravitational-Wave Observatories: Synergies with Other Scientific Communities} (11 pages) identifies a broad set of scientific constituencies beyond the ground-based GW community having common scientific interests where GW data have significant impact and makes a series of recommendations for facilitating strong linkages with those relevant scientific communities.  It also presents communication and outreach plans to help engage with those communities and motivate them in support of 3G GW science. 

\item \textbf{Gravitational-Wave Data Analysis: Computing Challenges in the 3G Era} (23 pages) lays out the many computational challenges -- required computing/storage/networking resources, anticipated algorithm and software needs, and commensurate person power -- imposed by the orders of magnitude greater GW event rates enabled by these new GW observatories.  Key to meeting these challenges will be establishing synergistic collaborations with the high energy physics and astronomy communities as well as with industrial partners to develop shared solutions.  

\item \textbf{An Exploration of Possible Governance Models for the Future Global Gravitational-Wave Observatory Network} (13 pages) evaluates a variety of possible governance models by carrying out a comprehensive survey of existing organizational structures of large-scale scientific laboratories and collaborations to propose a sustainable, centralized governance model for the overall management of the construction and operations of the planned 3G observatories.

\end{enumerate}

\section{Intended Audiences and Impacts}

These reports, somewhat technical in nature, are primarily intended for members of the GW community to serve as a basis for planning and executing successful proposals to construct and operate future 3G observatories. These reports are of particular relevance to two future ground-based observatory efforts -- the Einstein Telescope (ET) in Europe and the Cosmic Explorer (CE) in the US. ET and CE are in different stages of planning toward project approval at present, however both projects have had close interactions with the GWIC 3G Subcommittee during the course of the development of these reports and acknowledge benefiting from the material presented in them. 

The reports may also serve to familiarize and educate other individuals and groups, including researchers from other disciplines as well as funding agencies, on specific themes and topics of overlapping interest with the GW community.

Each report was written, to the extent possible, to be self-contained and stand-alone, however when read collectively they will provide a broader and more complete perspective on the scientific opportunities afforded by 3G observatories and the steps needed to realize their operation in the 2030s.

\newpage

\section{GWIC 3G Subcommittee Membership}\label{3gteam}

\textbf{Steering Committee}\\
Michele Punturo, INFN Perugia, Italy (Co-chair)\\
David Reitze, Caltech, USA (Co-chair)\\
Peter Couvares, Caltech, USA\\
Stavros Katsanevas, European Gravitational Observatory\\
Takaaki Kajita, University of Tokyo, Japan\\
Vicky Kalogera, Northwestern University, USA\\
Harald Lueck, AEI, Hannover, Germany\\
David McClelland, Australian National University, Australia\\
Sheila Rowan, University of Glasgow, UK\\
Gary Sanders, Caltech, USA\\
B.S. Sathyaprakash, Penn State University, USA and Cardiff University, UK\\
David Shoemaker, MIT, USA (Secretary)\\
Jo van den Brand, Nikhef, Netherlands\\

\textbf{Science Book Subcommittee}\\
Vicky Kalogera, Northwestern University, USA (Co-chair)\\
B.S. Sathyaprakash, Penn State USA and Cardiff University, UK (Co-chair)\\
Matthew Bailes, Swinburne, Australia\\
Marie Anne Bizouard, CNRS, France\\
Alessandra Buonanno, AEI, Potsdam, Germany, and University of Maryland, USA\\
Adam Burrows, Princeton, USA\\
Monica Colpi, INFN, Italy\\
Matt Evans, MIT, USA\\
Stephen Fairhurst, Cardiff University, UK\\
Stefan Hild, Maastricht University, Netherlands\\
Mansi M. Kasliwal, Caltech, USA\\
Luis Lehner, Perimeter Institute, Canada\\
Ilya Mandel, University of Birmingham, UK\\
Vuk Mandic, University of Minnesota, USA\\
Samaya Nissanke, University of Amsterdam, Netherlands\\
Marialessandra Papa, AEI, Hannover, Germany\\
Sanjay Reddy, University of Washington, USA\\
Stephan Rosswog, Oskar Klein Centre, Sweden\\
Chris Van Den Broeck, Nikhef, Netherlands\\

\textbf{Detector Research and Development Subcommittee}\\
David McClelland, Australian National University, Australia (Co-chair)\\
Harald Lueck, AEI, Hannover, Germany (Co-chair)\\
Rana Adhikari, Caltech, USA\\
Masaki Ando, University of Tokyo, Japan\\
GariLynn Billingsley, Caltech, USA\\
Geppo Cagnoli, ILM, Lyon, France\\
Matt Evans, MIT, USA \\
Martin Fejer, Stanford University, USA\\
Andreas Freise, University of Birmingham, UK\\
Paul Fulda, University of Florida, USA\\
Eric Genin, Virgo, Italy\\
Gabriela González, Louisiana State University, USA\\
Jan Harms, Universit\`a degli Studi di Urbino, Italy\\
Stefan Hild, University of Glasgow, UK\\
Giovanni Losurdo, INFN Pisa, Italy\\
Ian Martin, University of Glasgow, UK\\
Anil Prabhakar, IIT Madras, India\\
Stuart Reid, University of Strathclyde, UK\\
Fulvio Ricci, Universit\`a La Sapienza, and INFN Roma, Italy\\
Norna Robertson, Caltech, USA\\
Jo van den Brand, Nikhef, Netherlands\\
Benno Willke, AEI, Hannover, Germany\\
Mike Zucker, MIT, USA\\

\textbf{Synergies with Scientific Communities Subcommittee}\\
Michele Punturo, INFN Perugia, Italy (Co-chair)\\
David Reitze, Caltech, USA (Co-chair)\\
David Shoemaker, MIT, USA\\

\textbf{Governance Models Subcommittee}\\
Stavros Katsanevas, Astro Particle and Cosmology Laboratory, France (Co-chair)\\
Gary Sanders, Caltech, USA (Co-chair)\\
Beverly Berger, Stanford University, USA\\
Gabriela González, Louisiana State University, USA\\
James Hough, University of Glasgow, UK\\
Ajit K. Kembhavi, Inter-University Centre for Astronomy and Astrophysics, India\\
Frank Linde, Nikhef, Netherlands\\
David McClelland, Australian National University, Australia\\
Masatake Ohashi, Institute of Cosmic Ray Research, Japan\\
Fulvio Ricci, Universit\`a La Sapienza, and INFN Roma, Italy\\
Stan Whitcomb, Caltech, USA\\

\textbf{Data Analysis Computing Challenges Subcommittee}\\
Peter Couvares, Caltech, USA (Co-chair)\\
Ian Bird, CERN, Switzerland (Co-chair)\\
Ed Porter, Université Paris Diderot, France (Co-chair)\\
Stefano Bagnasco, INFN, Italy\\
Geoffrey Lovelace, California State Fullerton, USA\\
Josh Willis, Caltech, USA\\